\documentclass[a4paper, onecolumn, 10pt,accepted=2021-05-04]{quantumarticle}
 \pdfoutput=1 

\usepackage{amsfonts,amssymb,amsthm}
\usepackage{dsfont}
\usepackage[utf8x]{inputenc}

\usepackage{makeidx,graphics}
\usepackage{epsfig}
\usepackage{amsmath}
\usepackage{amscd}
\usepackage{longtable}
\DeclareGraphicsExtensions{bmp,eps}
\usepackage{graphicx}
\usepackage{braket}
\usepackage{color}

\usepackage[numbers,sort&compress]{natbib}

\usepackage[titletoc,title]{appendix}

\def\be{\begin{equation}}
\def\ee{\end{equation}}
\def\bea{\begin{eqnarray}}
\def\eea{\end{eqnarray}}

\newcommand{\sect}[1]{\setcounter{equation}{0}\section{#1}}
\newcommand{\subsect}[1]{\subsection{#1}}

\renewcommand{\theequation}{\arabic{section}.\arabic{equation}}

\begin{document}



\title{The group structure of dynamical transformations between quantum reference frames}

\author{Angel Ballesteros}
\affiliation{Departamento de F\'isica, Universidad de Burgos, 
09001 Burgos, Spain}
\email{angelb@ubu.es}

\author{Flaminia Giacomini}
\affiliation{Perimeter Institute for Theoretical Physics,
31 Caroline St. N, Waterloo, Ontario, N2L 2Y5, Canada}
\email{fgiacomini@perimeterinstitute.ca}

\author{Giulia Gubitosi}
\affiliation{Dipartimento di Fisica Ettore Pancini, Universit\`a di Napoli Federico II, and INFN, Sezione di Napoli, Complesso Univ. Monte S. Angelo, I-80126 Napoli, Italy}
\email{giulia.gubitosi@unina.it}

\maketitle

\begin{abstract}
\noindent
Recently, it was shown that when  reference frames are associated to  quantum systems, the transformation laws between such quantum reference frames need to be modified to take into account the quantum and dynamical features of the reference frames. This led to a relational description of the phase space variables of the quantum system of which the quantum reference frames are part of.
While such transformations were shown to be symmetries of the system's Hamiltonian, the question remained unanswered as to whether they enjoy a group structure, similar to that of the Galilei group relating classical reference frames in quantum mechanics.
In this work, we identify the canonical transformations on the phase space of the quantum systems comprising the quantum reference frames, and show that these transformations close a group structure defined by a Lie algebra, which is different from the usual Galilei algebra of quantum mechanics.  We further find that the elements of this new algebra are in fact the building blocks of the quantum reference frames transformations previously identified, which we recover.
Finally, we show how  the transformations between classical reference frames described by the standard Galilei group symmetries can be obtained from the group of transformations between quantum reference frames by taking the zero limit of the parameter that governs the additional noncommutativity introduced by the quantum nature of inertial transformations.

\end{abstract}
\medskip


\sect{Introduction}

In the standard description of quantum mechanics, the quantum states arising as solutions of the Schr\"odinger equation of a free particle are invariant under transformations between reference frames linked by (centrally extended) Galilei transformations~\cite{levy1971galilei,LevyLeblond:1974zj}. These transformations define a Lie group of  symmetries, such that the composition of any symmetry transformation results in another transformation belonging to the group. 
In this setting, while the system under study is quantum, the reference frames themselves are abstract entities with no quantum properties (they are sharply defined) and no dynamical behaviour.  They are usually identified with the ``laboratory'' within which the quantum system lives.
This dichotomy between the description of the system and that of the reference frames raises questions, since at least in some limit the quantum and dynamical properties of the reference frame might become non-negligible compared to those of the quantum system.

Associating reference frames to physical systems, which can ultimately be quantum systems, leads to the notion of Quantum Reference Frames (QRFs). QRFs have been extensively discussed in the literature in different contexts. In the quantum information literature \cite{aharonov1, aharonov2, aharonov3, brs_review, bartlett_communication, spekkens_resource, palmer_changing, bartlett_degradation, smith_quantumrf, poulin_dynamics, skotiniotis_frameness, angelo_1, angelo_2, angelo_3}, they have been shown to be a useful tool to overcome superselection rules, to devise communication tasks in the absence of a shared reference frame between two parties. Independently, some authors have argued that, in quantum gravity, the gauge nature of the gravitational field requires considering QRFs \cite{dewitt1967quantum, rovelli_quantum}, and that QRFs admit a description in terms of deformed symmetries \cite{poulin_deformed}. In addition, QRFs have been proposed to be relevant in a quantum formulation of the equivalence principle \cite{Hardy:2018kbp, hardy2020implementation, giacomini2020equivalence}. From a foundational perspective, quantum mechanics can be formulated in relational terms \cite{rovelli_relational} from the perspective of QRFs \cite{busch_relational_1, busch_relational_2, busch_relational_3, jacques}. Recently, a relational formulation of QRFs has been used to perform QRFs transformations in different contexts, such as Galilean \cite{QRF, perspective1, perspective2, yang2020switching} and special-relativistic \cite{giacomini2019relativistic,  streiter2020relativistic} quantum physics, quantum clocks models \cite{hoehn2018switch, hoehn2019trinity, castro2020quantum, hoehn2020equivalence}, cosmology \cite{hohn2019switching},  finite-dimensional quantum systems \cite{de2020quantum, krumm2020}, and superpositions of curved spacetimes \cite{giacomini2020equivalence}.
 
In this work, we are going to focus on the formalism introduced in Ref.~\cite{QRF}, where it was shown that one can develop a framework where the reference frames are part of the quantum system under study. In this setting, reference frames are assigned a possibly evolving quantum state and the phase space of the full quantum system is described via relational phase space coordinates. Of course, the laws of transformation between different QRFs must somewhat generalize  the usual Galilei transformations, in order to accommodate the quantum properties of the transformation parameters and the relational description of the phase space coordinates.
Thanks to these developments, the reference frames become less abstract entities, and can be identified with specific elements of the system. For example, one can consider a quantum system made up of several free particles, and describe it as it is seen by any one of them, and subsequently change the description to that corresponding to the point of view of a different particle.
The crucial point is that, when transforming the phase space coordinates from the point of view of one particle to the other, one ends up performing, in the terminology of Ref.~\cite{QRF}, an \textit{extended} Galilean transformation, where the commutative parameter of standard Galilean transformations is replaced by a quantum operator on the Hilbert space of the QRF. Moreover, since any of the particles can be taken as a QRF, it makes sense to describe the system in terms of relational phase space coordinates.

Going back to the standard description of quantum mechanics we illustrated above, two crucial features of the Galilei transformations between classical reference frames are that these transformations are in fact symmetries of the Hamiltonian of free particles and that they close a group, namely, the composition of any two Galilei transformations gives a transformation that is again a Galilei transformation. Of course in the context of the standard description of quantum mechanics the second feature is implied by the first one, but in the following we will treat the two properties separately. In fact, in the QRFs framework such implication is not obviously demonstrated and is the main subject of the work presented here.

Establishing whether QRF transformations describe some kind of symmetries of the quantum system is a non trivial task already. The question was answered positively in Ref.~\cite{QRF}, where it was demonstrated that QRF transformations can be constructed so to be extended symmetries of the Hamiltonian of the system.  As we will review in the following Sec.~\ref{sec:IntroQRF}, this result was achieved by realizing that the dynamical nature of QRFs requires   to include the time evolution operators explicitly in the QRF transformation, besides promoting the commutative parameters of standard Galilean transformations to quantum operators.
In this work, we take this result as a starting point and set to the task of investigating the group nature of QRF transformations.

In order to uncover the group structure of the transformations defined in Sec.~\ref{sec:IntroQRF}, in  Sec.~\ref{sec:FrozenAlg} we take one step back and look at the time-independent version of the QRF transformations corresponding to  generalised translation and boost transformations. Specifically, we consider the transformations in the form they take when the time evolution of the quantum system comprising the QRFs is ignored. The generators of these simplified transformations are identified with tensor product operators, acting on the tensor product of the phase space algebras of the QRF to which we are transforming and of the quantum particle which is being transformed. This allows us to find that these two operators, together with two additional ones, close a Lie algebra that will be called the {\it relational Lie algebra} for QRFs. Looking back at the results described in Sec.~\ref{sec:IntroQRF},  we observe that these two additional operators were already included in the full  QRF transformations for independent reasons. In fact, in the construction of  Sec.~\ref{sec:IntroQRF} the additional operators were needed to preserve the relational nature of the phase space coordinates, which would otherwise be lost if only acting with a generalised translation or boost. This fact motivates the underlying Lie symmetry to be called a relational one. So we find that, starting from just the   extended translation and boost  generators and  requiring that a group structure exists, we are forced to introduce all of the additional building blocks of the QRF transformations defined in Sec.~\ref{sec:IntroQRF}, and it is at the level of these building blocks that the group structure is apparent.

A similar result is found when the time evolution of the quantum system is accounted for, as we do in Sec.~\ref{sec:TimeDepAlg}. In this case we find that the Lie algebra of which the extended translation and boost generators are part is even larger, namely a 7-dimensional one, that will be called the {\it dynamical Lie algebra} for QRFs. Once more it only contains generators that had already been included in the QRF transformations of Sec.~\ref{sec:IntroQRF},  in order to ensure that the latter define extended symmetry transformations of the Hamiltonian and in order to preserve the relational nature of the phase space coordinates. The group structure we uncover for the building elements of the QRF transformations  is indeed reflected in the composition rule of  two QRF transformations, as we show in Sec.~\ref{sec:GroupLaw}. Finally, while the structures we build depart  significantly from the ones that are usual in quantum mechanics, in Sec.~\ref{sec:Classical} we are  able to expose the ``classical reference frame limiting procedure'' which allows us to recover the classical centrally extended Galilei algebra which describes the symmetries of a quantum particle in standard quantum mechanics.

Notice that a different study of the relation between QRFs and symmetry groups was carried out in Refs.~\cite{de2020quantum, krumm2020}. In particular, in Ref.~\cite{de2020quantum} it was shown that, within a formalism which identifies reference frames with elements of a symmetry group $G$, a consistent change of QRFs can be obtained if and only if the QRFs carry a regular representation of a symmetry group. In Ref.~\cite{krumm2020} it was shown that QRF transformations appear as symmetry transformations of the physical system where they are defined. These works deal with finite-dimensional systems, and focus on a different question ---understanding a specific QRF transformation as a symmetry of a given quantum system--- to the one we answer here, \emph{i.e.}, finding that there exists a Lie group of inertial transformations for QRFs that generalizes the Galilean inertial transformations for the corresponding classical reference frames.

\sect{Quantum reference frames transformations}
\label{sec:IntroQRF}

In this section we revisit the main results of Ref.~\cite{QRF}, introducing a slightly different notation than the one used in the original work, which is more suitable for the algebraic analysis we carry in the following sections. In addition, we emphasise some conceptual aspects of the QRF formalism, such as the way a measurement is described from the point of view of two different QRFs, in order to make the presentation self-contained.

 A QRF corresponds to a physical system to which a set of coordinates is associated, and which can be in a quantum relationship with other physical systems. This formalism is operational, in that primitive laboratory operations ---preparation, transformations, and measurements of quantum states--- have fundamental status, and relational, because everything is formulated in terms of relational quantities and the formalism does not require the presence of any external or absolute reference frame. 

In order to illustrate the idea of QRFs, we consider the simple situation of a  quantum system composed of three free particles. We describe quantum particles $A$ and $B$ in terms of the relative coordinates to the initial QRF identified with particle $C$. The relational phase space variables $\{\hat x_A^{(C)}, \hat p_A^{(C)}, \hat x_B^{(C)}, \hat p_B^{(C)}\}$ are given, respectively, by the position and momenta of particles $A$ and $B$ as seen from $C$. We use the notation $x_{i}^{(j)}$ to identify the coordinate of system $i$ as seen by reference $j$, and the analogous notation $p_{i}^{(j)}$ holds for momenta. Throughout the paper we work in 1 spatial dimension. Differently to Ref.~\cite{QRF}, we take two different 3-dimensional Heisenberg-Weyl algebras $h(3)$~\cite{Wolf} to distinguish the effects due to the quantum behaviour of particles $A$ and $B$:
\begin{equation}
[\hat x_A^{(C)},\hat p_A^{(C)}]=i\,\kappa \,,\qquad\qquad
[\hat x_B^{(C)},\hat p_B^{(C)}]=i\,\hbar \, ,
\label{commkh}
\end{equation}
where we use $\hbar$ for $B$, \emph{i.e.} the quantum system  that is being transformed, which has  the usual properties of the standard description of quantum mechanics, while we use a different constant $\kappa$ for $A$, since this is taken as a QRF, while in usual quantum mechanics this would be a classical object. 
We will see that a consistent treatment of QRF transformations requires $\kappa=\hbar$, since $A$ and $B$ can alternate in playing the role of observed quantum system and QRF, but keeping the two constants formally different  allows us to track the effects due to the quantum and dynamical nature of the reference frame to which we transform, as well as to perform the appropriate smooth limit to a classical reference frame scenario, which we do in Sec.~\ref{sec:Classical}.

\subsection{QRF transformations and expectation values}

The QRF transformation is a unitary transformation $\hat{S}$ which maps the quantum state in the Hilbert space of $A$ and $B$ relative to C, $\mathcal{H}_A^{(C)} \otimes \mathcal{H}_B^{(C)}$, to the quantum state in the Hilbert space of $B$ and $C$ relative to $A$, \emph{i.e.}, $\mathcal{H}_B^{(A)} \otimes \mathcal{H}_C^{(A)}$.\footnote{Notice that one can perform a completely analogous transformation by taking $B$ instead of $A$ as the new reference frame. Likewise, this symmetric role of $A$ and $B$ applies  to all the transformations we describe in the following.} The unitary action on a quantum state is then $\hat{S}^{(C \rightarrow A)}\ket{\psi}_{AB}^{(C)} = \ket{\psi}_{CB}^{(A)}$. 

The property that the outcome of an experiment should be consistent in different (quantum) reference frames is encoded in the conservation of probabilities. In particular, the probability of detecting an outcome $b^*$ by measuring an observable $\hat{O}_{AB}^{(C)}$ in the initial QRF $C$ is 
\begin{equation}
	p(b^*)=\text{Tr}\left[ \hat{\rho}_{AB}^{(C)} \hat{O}_{AB}^{(C)}(b^*) \right],
\end{equation}
where $\hat{\rho}_{AB}^{(C)}$ is the quantum state from the perspective of $C$ and $\hat{O}_{AB}^{(C)}(b^*)$ is the projector on the outcome $b^*$. The conservation of probabilities follows immediately from the unitarity of the QRF transformation
\begin{equation} \label{eq:MeasA}
	p(b^*)=\text{Tr}\left[ \hat{\rho}_{BC}^{(A)} \hat{O}_{BC}^{(A)}(b^*) \right],
\end{equation} 
where $\hat{\rho}_{BC}^{(A)} = \hat{S}^{(C\rightarrow A)} \hat{\rho}_{AB}^{(C)} (\hat{S}^{(C\rightarrow A)})^\dagger$ and $\hat{O}_{BC}^{(A)} = \hat{S}^{(C\rightarrow A)} \hat{O}_{AB}^{(C)} (\hat{S}^{(C\rightarrow A)})^\dagger$. Clearly, this result applies also for expectation values of observables.

The simplest transformation to relative positions $\hat{S}_x^{(C \rightarrow A)}$ is defined as
\begin{equation}\label{eq:translation}
	\hat{S}_x^{(C \rightarrow A)} = \mathcal{\hat{P}}_{AC} e^{\frac{i}{\hbar}\hat{x}_A^{(C)} \otimes \hat{p}_B^{(C)}},
\end{equation} 
where $\hat{x}_i^{(C)}, \hat{p}_i^{(C)}$, with $i=A,B$, are the phase space operators defined above, and $\mathcal{\hat{P}}_{AC}$ is an operator, named ``parity-swap'' operator, which acts on $A$ as 
\begin{equation}
\mathcal{\hat{P}}_{AC} \hat{x}_A^{(C)} \mathcal{\hat{P}}_{AC}^\dagger = - \hat{x}_C^{(A)} ,\qquad
\mathcal{\hat{P}}_{AC} \hat{p}_A^{(C)} \mathcal{\hat{P}}_{AC}^\dagger = - \hat{p}_C^{(A)} ,
\end{equation}
and on $B$ as
\begin{equation}
\mathcal{\hat{P}}_{AC} \hat{x}_B^{(C)} \mathcal{\hat{P}}_{AC}^\dagger =  \hat{x}_B^{(A)},\qquad
\mathcal{\hat{P}}_{AC} \hat{p}_B^{(C)} \mathcal{\hat{P}}_{AC}^\dagger =  \hat{p}_B^{(A)}.
\end{equation}
The presence of such parity-swap operator is required in order to map consistently the full set of relative positions of $A$ and $B$ from the point of view of $C$ to the relative position of $B$ and $C$ from the point of view of A. Explicitly, the QRF transformation $\hat{S}_x^{(C \rightarrow A)}$ acts as (we leave the indication ${(C \rightarrow A)}$ implicit)
\begin{equation}
	\begin{split}
		\hat{S}_x \hat{x}_A^{(C)} \hat{S}_x^\dagger = -\hat{x}_C^{(A)}; &\qquad \hat{S}_x \hat{p}_A^{(C)} \hat{S}_x^\dagger = - \hat{p}_C^{(A)} -\frac{\kappa}{\hbar}\hat{p}_B^{(A)};\\
		\hat{S}_x \hat{x}_B^{(C)} \hat{S}_x^\dagger = \hat{x}_B^{(A)} - \hat{x}_C^{(A)}; &\qquad \hat{S}_x \hat{p}_B^{(C)} \hat{S}_x^\dagger = \hat{p}_B^{(A)}.
	\end{split}
\end{equation}
This transformation maps the position operators of $A$ and $B$ relative to $C$, i.e., $\hat{x}_A^{(C)}$ and $\hat{x}_B^{(C)}$, into the position operators of $B$ and $C$ relative to $A$, i.e.,  $\hat{x}_B^{(A)}$ and $\hat{x}_C^{(A)}$, and the momenta  $\hat{p}_A^{(C)}$ and $\hat{p}_B^{(C)}$ into the canonically conjugated operators to $\hat{x}_B^{(A)}$ and $\hat{x}_C^{(A)}$, respectively $\hat{p}_B^{(A)}$ and $\hat{p}_C^{(A)}$. 

This can be easily seen by describing what a position measurement would look like as described from the perpective of QRF $C$ and $A$. In particular, let us consider a simple (non-normalised) state of $A$ and $B$ in the initial QRF $C$, i.e., $\ket{a}_A \ket{b}_B$, where $a$ and $b$ are respectively the position of systems $A$ and $B$ relative to $C$. The expectation value of a measurement of the position of $B$ relative to $C$ is then
\begin{equation}
	\langle \hat{x}_B^{(C)} \rangle=\text{Tr}\left[ \hat{\rho}_{AB}^{(C)} \hat{x}_{B}^{(C)} \right] = {}_B\bra{b} {}_A\bra{a} \hat{x}_B^{(C)} \ket{a}_A \ket{b}_B =b,
\end{equation}
provided that we correctly normalise the quantum state. When we describe this measurement from the point of view of $A$, we need to transform both the state, which becomes $\hat{S}_x \ket{a}_A \ket{b}_B = \ket{b-a}_B \ket{-a}_C$, and the observables. By doing this, we obtain the same expectation value, but we describe this experiment as a measurement on both systems $B$ and $C$, and specifically as a measurement of their relative position, i.e.,
\begin{equation}
	\langle \hat{x}_B^{(A)} - \hat{x}_C^{(A)} \rangle=\text{Tr}\left[ \hat{\rho}_{BC}^{(A)} (\hat{x}_B^{(A)} - \hat{x}_C^{(A)})\right] = {}_C\bra{-a} {}_B\bra{b-a} (\hat{x}_B^{(A)} - \hat{x}_C^{(A)}) \ket{b-a}_B \ket{-a}_C= b,
\end{equation}
as prescribed by Eq.~\eqref{eq:MeasA}. Notice that it is perfectly admissible to measure $\hat{x}_B^{(A)}$ in the QRF of $A$, but it would correspond to a different experiment to the one described above, equivalent to measuring the observable $\hat{x}_B^{(C)} - \hat{x}_A^{(C)}$ in the initial QRF $C$, and would give the consistent prediction, in both QRFs, that $\langle \hat{x}_B^{(C)} - \hat{x}_A^{(C)} \rangle = \langle \hat{x}_B^{(A)} \rangle = b-a$. The treatment of measurements in different QRFs shows that the transformation $\hat{S}_x$ preserves the relational character of the transformation, because each position operator in different QRFs correspond to the measurement of the relative distance of the system they refer to and the origin of the QRF. In addition, the measurement procedure, as explained above, gives consistent predictions in the initial and final QRF. The full description of the measurement in different QRFs is detailed in the \emph{Methods - Measurements as seen from a quantum reference frame} in Ref.~\cite{QRF}.

\subsection{Extended symmetry transformations}

An especially interesting class of QRF transformations is the one that achieves an \textit{extended symmetry transformation} for the free-particle Hamiltonian. An \textit{extended symmetry transformation} was defined in Ref.~\cite{QRF} as a QRF tranformation mapping a Hamiltonian in the initial QRF $C$ to a Hamiltonian in the final QRF $A$ having the same functional form as the initial Hamiltonian, but with all labels $A$ and $C$ interchanged. In general, given the initial Hamiltonian $\hat{H}_{AB}^{(C)}$ of $A$ and $B$ from the point of view of C, the Hamiltonian in the QRF $A$ is obtained via a QRF transformation $\hat{S}^{(C \rightarrow A)}$ as
\begin{equation}
	\hat{H}_{BC}^{(A)} = \hat{S} \hat{H}_{AB}^{(C)} \hat{S}^\dagger + i\hbar \frac{d \hat{S}}{dt}\hat{S}^\dagger.  \label{eq:ExtSymm}
\end{equation} 
A QRF transformation is an extended symmetry transformation if it maps the initial Hamiltonian $\hat{H}_{AB}^{(C)} = \frac{(\hat{p}_A^{(C)})^2}{2m_A} + \frac{(\hat{p}_B^{(C)})^2}{2m_B}$ to $\hat{H}_{BC}^{(A)} = \frac{(\hat{p}_B^{(A)})^2}{2m_B} + \frac{(\hat{p}_C^{(A)})^2}{2m_C}$. Such extended symmetry transformations constitute the inertial transformations for QRFs corresponding to the \textit{superposition of Galilean translations} and the \textit{superposition of Galilean boosts}, respectively
\begin{align}
	&\hat{S}_T^{(C \rightarrow A)} = e^{-\frac{i}{\kappa}\hat{Q}_C^{(A)} t} \mathcal{\hat{P}}_{AC} e^{\frac{i}{\hbar} \hat{P}_{AB}^{(C)}}  e^{\frac{i}{\kappa}\hat{Q}_A^{(C)} t}, \label{eq:extT}\\
	&\hat{S}_b^{(C \rightarrow A)} = e^{-\frac{i}{\kappa}\hat{Q}_C^{(A)} t} \mathcal{\hat{P}}_{AC} e^{\frac{i}{\kappa} \ln \left(\frac{m_C}{m_A}\right) \hat{D}_A^{(C)}} e^{\frac{i}{\hbar} \hat{K}_{AB}^{(C)}}  e^{\frac{i}{\kappa}\hat{Q}_A^{(C)} t},\label{eq:extb}
\end{align}
where $\hat{Q}_i^{(j)} = \frac{(\hat{p}_i^{(j)})^2}{2m_i}$ is the free-particle Hamiltonian of $i$ from the point of view of $j$, and with
\begin{align}
	& \hat{P}_{AB}^{(C)} = \hat{x}_A^{(C)} \otimes \hat{p}_B^{(C)},\\
	& \hat{D}_A^{(C)} = \frac{1}{2}\left(\hat x_A^{(C)}\,\hat p_A^{(C)} + \hat p_A^{(C)}\,\hat x_A^{(C)} \right),\\
	& \hat K_{AB}^{(C)}= \frac{\hat p_A^{(C)}}{m_A}\,\otimes \hat G_B^{(C)},
\end{align}
where $ \hat{G}_B^{(C)} = \hat{p}_B^{(C)} t -m_B \hat{x}_B^{(C)}$ is the generator of the standard Galilean boost on particle B. Compared to the definition of $\hat{S}_T^{(C\rightarrow A)}$ and $\hat{S}_b^{(C\rightarrow A)}$ given in Ref.~\cite{QRF}, here we have distinguished the two constants $\hbar$ and $\kappa$. We use them, respectively, when the transformation acts on $B$ or exclusively on $A$. In this way $\kappa$ will highlight all symmetry operations defined on the A phase space only, although indeed the transformations \eqref{eq:extT} and \eqref{eq:extb} are symmetries only if $\hbar=\kappa$. Of course the analogous transformations $\hat{S}_T^{(C \rightarrow B)}$ and $\hat{S}_b^{(C \rightarrow B)}$ are  extended symmetries of the Hamiltonian as well for $\hbar=\kappa$. Notice that the left-most term in Eqs.~\eqref{eq:extT}, \eqref{eq:extb} can be commuted with the parity-swap operator and equivalently written as
\begin{equation}
	e^{-\frac{i}{\kappa}\hat{Q}_C^{(A)} t} \mathcal{\hat{P}}_{AC} = \mathcal{\hat{P}}_{AC} e^{-\frac{i}{\kappa}\frac{m_A}{m_C}\hat{Q}_A^{(C)} t}.
\end{equation}
  The transformations of Eqs.~\eqref{eq:extT}-\eqref{eq:extb} are generalizations of the standard translation and Galilean boost transformations respectively, in a sense that will be explained shortly. Besides including quantum transformation parameters and the parity swap operators\footnote{Notice that for the boost transformation the full parity swap operator which preserves the relational nature of the phase space coordinates is $ \mathcal{\hat{P}}_{AC} e^{\frac{i}{\kappa} \ln \left(\frac{m_C}{m_A}\right) \hat{D}_A^{(C)}}$ so to also exchange appropriately the velocities of particles $A$ and $C$.}, as was done e.g. for the quantum translation transformation of Eq.~\eqref{eq:translation}, the requirement that these transformations are extended symmetry transformations leads to further including the time evolution operators of the systems $A$ and C, $e^{\pm\frac{i}{\kappa}\hat{Q}_i^{(j)} t}$.   We stress that, consistently with Eq.~\eqref{commkh}, elementary unitary transformations acting on A(C) are ruled by the $\kappa$ parameter, while the ones acting on particle $B$ contain $\hbar$. 

The transformation $\hat{S}_T^{(C \rightarrow A)}$ corresponding to the ``superposition of Galilean translations'' acts as
\begin{equation}
	\begin{split}
		&\hat{S}_T \hat{x}_A^{(C)} \hat{S}_T^\dagger = -\hat{x}_C^{(A)} + \hat{p}_C^{(A)} t\left[\frac{1}{m_C}-\frac{1}{m_{A}}\right]- \frac{\kappa}{\hbar}\frac{ \hat{p}_B^{(A)}}{m_A}t,\\
		& \hat{S}_T \hat{p}_A^{(C)} \hat{S}_T^\dagger =  - \hat{p}_C^{(A)} -\frac{\kappa}{\hbar}\hat{p}_B^{(A)},\\
		&\hat{S}_T \hat{x}_B^{(C)} \hat{S}_T^\dagger = \hat{x}_B^{(A)} - \hat{x}_C^{(A)} + \frac{\hat{p}_C^{(A)}}{m_C}t,\\
		& \hat{S}_T \hat{p}_B^{(C)} \hat{S}_T^\dagger = \hat{p}_B^{(A)}.
	\end{split}\label{eq:STaction}
\end{equation}
The physical meaning of this transformation is that the position of system $B$ at time $t$ from the point of view of $C$ is mapped to the relative position between system $B$ at time $t$ and system $C$ at time $t=0$, while the momentum of $B$ remains unchanged (as we expect for a translation). If $\hbar=\kappa$ this construction of the QRF transformation satisfies the transitive property, meaning that the change of reference frame from $C$ to $A$ $\hat{S}_T^{(C \rightarrow A)}$ is equivalent to the change of reference frame from $C$ to $B$ and then from $B$ to A, \emph{i.e.}, $\hat{S}_T^{(C \rightarrow A)} = \hat{S}_T^{(B \rightarrow A)} \hat{S}_T^{(C \rightarrow B)}$.  The fact that transitivity requires $\hbar=\kappa$ should not come as a surprise, since when composing the two transformations the systems $A$ and $B$ play alternatively the role of transformed quantum system and of QRF, so that their treatment should be symmetric. Notice that, when $\kappa \rightarrow 0$, the action of the transformation of system $B$ is the same as with finite $\kappa$, but the action of the transformation on the QRF $A$ becomes independent of the position and momentum of system B.

The transformation $\hat{S}_b^{(C \rightarrow A)}$ corresponds  to the ``superposition of Galilean boosts'', and acts as
\begin{equation}
	\begin{split}
		&\hat{S}_b \hat{x}_A^{(C)} \hat{S}_b^\dagger = -\frac{m_C}{m_A}\hat{x}_C^{(A)} +\hat{p}_C^{(A)}t \left[\frac{1}{m_A} - \frac{1}{m_C} \right] + \frac{\kappa}{\hbar} \,\frac{\hat{p}_B^{(A)} t-m_{B}  \hat{x}_B^{(A)}}{m_A},\\
		& \hat{S}_b \hat{p}_A^{(C)} \hat{S}_b^\dagger = - \frac{m_A}{m_C}\hat{p}_C^{(A)},\\
		&\hat{S}_b \hat{x}_B^{(C)} \hat{S}_b^\dagger = \hat{x}_B^{(A)} -  \frac{\hat{p}_C^{(A)}}{m_C}t,\\
		& \hat{S}_b \hat{p}_B^{(C)} \hat{S}_b^\dagger = \hat{p}_B^{(A)} - \frac{m_B}{m_C}\hat{p}_C^{(A)}.
	\end{split}\label{eq:Sbaction}
\end{equation}
In Appendix~\ref{App:CalcD} we provide an explicit calculation of the action of the operator $e^{\frac{i}{\hbar}\ln \frac{m_C}{m_A} \hat{D}_A}$ on the phase space operators. The physical meaning of this transformation is that system $B$ at time $t$ from the point of view of $C$ is boosted  by an amount \textit{controlled} by the momentum (velocity) of system A. When $\kappa \rightarrow 0$, the action of the transformation of system $B$ is the same as with finite $\kappa$, but the action of the transformation on the QRF $A$ becomes independent of the position and momentum of system B.
Similarly to the previous case, if $\hbar=\kappa$ also this QRF transformation satisfies the transitive property, \emph{i.e.}, $\hat{S}_b^{(C \rightarrow A)} = \hat{S}_b^{(B \rightarrow A)} \hat{S}_b^{(C \rightarrow B)}$ \cite{QRF}.  

A natural question at this point is what happens when two different extended symmetry transformations are composed, for  instance $\hat{S}_D^{(C \rightarrow A)}=\hat{S}_b^{(B \rightarrow A)} \hat{S}_T^{(C \rightarrow B)}$, where we take $\hbar=\kappa$ as explained above. One would expect that the extended symmetry transformations close some group structure, analogously to what happens for the standard Galilei transformations. However,  the group structure of Galilei transformations relies crucially on the classical nature of the transformation parameters, whose addition rule is given by the group structure of the transformations \cite{levy1971galilei}. In the case of QRF transformations matters become complicated due to the quantum nature of  the transformation parameters. In the rest of the paper we show how we can recover a group structure for QRF transformations, and how the transformation $\hat{S}_D^{(C \rightarrow A)}$ is written as a function of the elements of the group that we derive.

\sect{Time-independent transformations: the relational Lie algebra}
\label{sec:FrozenAlg}

In order to investigate the group structure of the QRF transformations we start from a simplified scenario, where we ignore the time evolution of the quantum system of which the reference frames are part.

Considering the quantum system described in the previous section, comprising quantum particles $A, B, C$, we aim at describing the transformation from the QRF $C$ to $A$. We start with the two Heisenberg-Weyl algebras of Eq.~\eqref{commkh}:
\be
[\hat x_A,\hat p_A]=i\,\kappa \,,\qquad\qquad
[\hat x_B,\hat p_B]=i\,\hbar \, .
\ee
where we omitted the ${(C)}$ index for simplicity.
When ignoring the dynamics of the quantum particles, we can generalise the operators corresponding to the translations and Galilean boosts acting on the particle $B$. This is done by exponentiating the standard Galilei generators at time $t=0$, namely the generator of translations  $\hat P_B=\hat p_B$ and  the boost generator $
\hat G_B= -m_B \hat{x}_B$, and introducing non-commuting transformation parameters, $\hat x_A$ for translations and $\hat p_A/m_{A}$ for boosts, which encode the quantum properties of the reference frame one is transforming to. Thus we define
\begin{equation} \label{tyb}
	\hat U_P={\rm e}^{\frac{i}{\hbar} \hat x_A\otimes \hat P_B}, \qquad\qquad
\hat U_G={\rm e}^{\frac{i}{\hbar} \frac{\hat p_A}{m_A}\,\otimes \hat G_B} \,.
\end{equation}
The $\otimes$ symbol is included in order to emphasise the fact that in the QRFs setting such ``quantum Galilei'' transformations are defined within the tensor space of two different non-Abelian algebras acting on different spaces.  The algebra $A$ can be interpreted as the algebra generated by the non-commuting translation $\hat x_A$ and boost $\hat p_A$ parameters of such ``quantum Galilei'' transformations. On the other hand, the Heisenberg-Weyl algebra $B$ is the usual quantum mechanical algebra of position and momentum operators for the particle B, from which the generators of the  Galilei algebra of usual inertial transformations are constructed. This algebraic framework is inspired by the theory of quantum groups, which are generalizations of Lie groups where  group parameters become noncommutative operators~\cite{Chari:1994pz}.

Because of  the noncommutativity of the transformation parameters, the operators $\hat U_P$ and $\hat U_G$ act on  the phase space  of both particles $A$ and $B$. The generalised translation operator acts as follows:
\begin{align} 
	\begin{split}\label{eq:UPaction}
		&\hat {U}_P \, \hat{x}_A \, \hat {U}_P^{-1}= \hat{x}_A,
\\
	&\hat {U}_P \, \hat{p}_A \, \hat {U}_P^{-1}=\hat{p}_A - \frac{\kappa}{\hbar}\, \hat{p}_B,
\\
	& \hat {U}_P \, \hat{x}_B \, \hat {U}_P^{-1}=   \hat{x}_B + \hat{x}_A, 
\\
	&\hat {U}_P \, \hat{p}_B \, \hat {U}_P^{-1}=\hat{p}_B,
	\end{split}
\intertext{while the generalised boost transformations act as}
	\begin{split} \label{eq:UGaction}
		& \hat {U}_G \, \hat{x}_A \, \hat {U}_G^{-1}= \hat{x}_A + \frac{\kappa}{\hbar}\, \frac{\hat{G}_B}{m_A} =
\hat{x}_A - \frac{\kappa}{\hbar}\, \frac{m_B}{m_A}\, \hat{x}_B,\\
		&  \hat {U}_G \, \hat{p}_A \, \hat {U}_G^{-1}=\hat{p}_A ,\\
		&\hat {U}_G \, \hat{x}_B \, \hat {U}_G^{-1}=   \hat{x}_B,\\
		&  \hat {U}_G \, \hat{p}_B \, \hat {U}_G^{-1}=\hat{p}_B + \frac{m_B}{m_A}\,\hat{p}_A. 
	\end{split}
\end{align}

From these results we see that the action of the quantum translation operator $\hat {U}_P$ on the particle $B$ is the same as we would expect from the usual  action of a Galilei translation, since $\hat{x}_B$ is translated by $\hat{x}_A$ and momentum $\hat{p}_B$ is not changed. Additionally, $\hat {U}_P$ has a  non-trivial action  on the momentum of the $A$ particle. Similarly, the boost operator $\hat U_{G}$ has the same action on the particle $B$ as the  one expected from the standard Galilei boost, while it has a non-trivial action on particle $A$, whose position operator is modified in terms of the position operator for $B$. We see that, as expected, the nontrivial features of these transformations are governed by the parameter $\kappa$, which sets the noncommutativity of the phase space of the  QRF $A$.

Going back to the QRF transformations of the previous section, notice that the transformations defined by $\hat U_{P}$ and $\hat U_{G}$ in Eq.~\eqref{tyb}, correspond to the extended symmetry transformations in Eqs.~\eqref{eq:extT}-\eqref{eq:extb} taken at time $t=0$ and up to the (generalised, in the case of boost transformations) parity swap operators. This is also reflected in the explicit form of the action on phase space coordinates, compare Eqs.~\eqref{eq:STaction}-\eqref{eq:Sbaction} to Eqs.~\eqref{eq:UPaction}-\eqref{eq:UGaction}. So at this level the transformations $\hat U_{P}$ and $\hat U_{G}$ can be seen as   QRF transformations which spoil the relational nature of the phase space coordinates.  

As we mentioned at the beginning of this section, the reason for introducing these transformations is to investigate the group structure of QRF transformations in a scenario that is stripped of all non-essential structure, with the  hope of learning something that can be usefully applied also to the more refined framework of Sec.~\ref{sec:IntroQRF}. Working with the operators $U_{P}$ and $U_{G}$, the question of whether they belong to some group structure can be investigated by searching for the corresponding algebra for  the operators $ \hat P_{AB}= \hat x_A\otimes \hat p_B$ and $\hat K_{AB}= \frac{\hat p_A}{m_A}\,\otimes \hat G_B$. The main difficulty, compared to the usual Galilei group, is that we are dealing with tensor product operators acting of the phase spaces of both $A$ and $B$ rather than the usual translation and boost operators acting on the phase space of $B$ alone.
Nevertheless, we are able to find the sought for algebra, which includes two additional generators besides the ones we started from. In fact, it is a matter of straightforward computation to show that the tensor product operators
\begin{equation} \label{eq:nondyngenerators}
	\begin{split}
		& \hat P_{AB}= \hat x_A\otimes \hat p_B, \\
		& \hat K_{AB}= \frac{\hat p_A}{m_A}\,\otimes \hat G_B, \\
		& \hat D_A= \frac{1}{2}\left(\hat x_A\,\hat p_A + \hat p_A\,\hat x_A \right)\,\otimes \mathds{1}_B, \\
		& \hat D_B= \mathds{1}_A \, \otimes \frac{1}{2} \left(\hat x_B \, \hat p_B + \hat p_B \, \hat x_B \right) 
	\end{split}
\end{equation}
satisfy the following commutation rules 
\begin{equation}  \label{eq:RelationalAlgebra}
	\begin{split}
		& [\hat K_{AB},\hat P_{AB}]=i \kappa\, \frac{m_B}{m_A}\,\hat D_B - i \hbar\, \frac{m_B}{m_A}\,\hat D_A, \\
		& [\hat D_A,\hat P_{AB}]=-\,i\,\kappa \,\hat P_{AB}, \\
		& [\hat D_B,\hat P_{AB}]=\,i\,\hbar \,\hat P_{AB}, \\
		& [\hat D_A,\hat K_{AB}]= \,i\,\kappa \,\hat K_{AB}, \\
		& [\hat D_B,\hat K_{AB}]= -\,i\,\hbar \,\hat K_{AB}, \\
		& [\hat D_A,\hat D_B]= 0.
	\end{split}
\end{equation}
Since these are \textit{linear} commutators, the four operators~\eqref{eq:nondyngenerators} generate a 4-dimensional Lie algebra and we will call Eq.~\eqref{eq:RelationalAlgebra} the {\it relational Lie algebra}  $\mathcal{R}(4)$ for QRFs. The reason for this terminology resides in the physical interpretation of the operators $\hat D_A$, $\hat D_B$. 
The operator $\hat x_i\,\hat p_i + \hat p_i\,\hat x_i$ for $i=A, B$ acts on the $\hat x_i$ and $\hat p_i$ operators by performing a dilation in each quadrature. More precisely, the action of this operator on the position and momentum operators of the the two quantum systems is (see also Appendix~\ref{App:CalcD})
\begin{equation} \label{eq:DADB}
	\begin{split} 
		{\rm e}^{\frac{i}{\kappa}\alpha \hat D_A} \hat x_A {\rm e}^{-\frac{i}{\kappa}\alpha \hat D_A} = {\rm e}^{ \alpha} \hat x_A, &\qquad {\rm e}^{\frac{i}{\kappa}\alpha \hat D_A} \hat p_A {\rm e}^{-\frac{i}{\kappa}\alpha \hat D_A} = {\rm e}^{- \alpha} \hat p_A,\\
		{\rm e}^{\frac{i}{\hbar}\beta \hat D_B} \hat x_B {\rm e}^{-\frac{i}{\hbar}\beta \hat D_B} = {\rm e}^{\beta} \hat x_B, &\qquad {\rm e}^{\frac{i}{\hbar}\beta \hat D_B} \hat p_B {\rm e}^{-\frac{i}{\hbar}\beta \hat D_B} = {\rm e}^{-\beta} \hat p_B.
	\end{split}
\end{equation}
Hence, the role of this operator is to rescale the position and momentum operators of the two quantum systems.  An operator of this kind is exactly the one involved in the ``generalised'' parity swap operator that enters in the QRF boost transformation of Eq.~\eqref{eq:extb}, and guarantees that the phase space transformation defined by $\hat S_{b}$ preserves the relational nature of the phase space coordinates. In the case of Eq.~\eqref{eq:extb} the operator is used to enforce the condition that the velocity of the final QRF $A$, as seen from the initial QRF $C$, is equal to the opposite of the velocity of $C$ as seen from $A$. Mathematically, this condition is written as
\be
	\hat p_A \mapsto - \frac{m_A}{m_C}\hat p_C.
\ee
It is clear that, by choosing $\alpha =  \ln \frac{m_C}{m_A}$ in Eq.~\eqref{eq:DADB}, and by subsequently applying the parity-swap operator $ \mathcal{\hat P}_{AC}$ introduced in Sec.~\ref{sec:IntroQRF}, this condition can be implemented.

So we have taken a first  meaningful step towards the understanding of the group structure of the QRF transformations discussed in Sec.~\ref{sec:IntroQRF}. Not only have we showed that the simplified QRF transformations $\hat U_{P}$ and $\hat U_{G}$ form a group. By requiring that they do so we are  automatically led to solve the issue that $\hat U_{P}$ and $\hat U_{G}$ break the relational meaning of the phase space coordinates $\hat x_{i}$, $\hat p_{j}$. In fact,  in order to close the algebra of generators we need to introduce the  velocity rescaling operators $\hat D_{A}$, $\hat D_{B}$ that are indeed the ones enforcing that the velocity of $A$ as seen from $C$ is the opposite of the velocity of $C$ as seen from A. This condition thus guarantees that the relational quantities are mapped consistently and symmetrically under QRF transformation.  

Before proceeding to the following section, in which we expand the scope of our analysis by restoring the dynamical properties of the QRFs, we make one final observation on the properties of the algebra of Eq.~\eqref{eq:RelationalAlgebra}. By merging the two generators $\hat D_A$ and $\hat D_B$ into one single generator $\hat D$ 
\begin{equation} \label{eq:singleDgen}
	\hat D= \kappa\, \frac{\mathds{1}_A}{m_A}\otimes m_B \hat D_B - 
\hbar \, \frac{\mathds{1}_A}{ m_A}\,\hat D_A\otimes m_B \mathds{1}_B,
\end{equation}
we get a three-dimensional subalgebra
\begin{equation} \label{su11}
	\begin{split}
		& [\hat P_{AB},\hat K_{AB}]=-i\,\hat D, \\
		& [\hat P_{AB},\hat D]=-2\,i\,\kappa\,\hbar\,\frac{m_B}{m_A} \,\hat P_{AB}, \\
		& [\hat K_{AB},\hat D]=  2\,i\,\kappa\,\hbar\,\frac{m_B}{m_A} \,\hat K_{AB} ,
	\end{split}
\end{equation}
which is just the $so(2,1)\simeq sl(2,\mathbb R)\simeq su(1,1)$ real Lie algebra (see~\cite{Wolf,Gilmore}). The $\kappa\to 0$ limit of this algebra  is a 3-dimensional Heisenberg-Weyl algebra $h(3)$, which is indeed generated by $\hat G_B,\hat p_B$ and $m_B\,\mathds{1}_B$. Note also that the generator $\hat D^\ast=\kappa\, \frac{\mathds{1}_A}{m_A}\otimes m_B \hat D_B +
\hbar \, \frac{\mathds{1}_A}{ m_A}\,\hat D_A\otimes m_B \mathds{1}_B$ is such that $[\hat D^\ast,\hat P_{AB}]=[\hat D^\ast,\hat K_{AB}]=[\hat D^\ast,\hat D]=0$, and the relational Lie algebra $\mathcal{R}(4)$ is thus  isomorphic to the direct sum of $so(2,1)$ with a central extension generated by $D^\ast$.

\sect{Time-dependent transformations: the dynamical Lie algebra}
\label{sec:TimeDepAlg}

As we have seen in Sec.~\ref{sec:IntroQRF}, the extended symmetry transformations for QRFs include information about the dynamical behaviour of the quantum system. This is provided by a Hamiltonian $\hat H$ describing the free relative motion of the quantum particles $A$ and $B$, namely
\be
\hat H=\frac{\hat p_A^2}{2\,m_A}\otimes I + I \otimes \frac{\hat p_B^2}{2\,m_B} \, .
\ee
In order to analyse in algebraic terms such dynamical content in a similar way as done in the previous section, we modify the generator $\hat K_{AB}$ appearing in the non-dynamical algebra of Eq.~\eqref{eq:nondyngenerators} by considering the time-dependent version of the Galilean boost, which amounts to replacing the operator $\hat G_B$ defined in the previous section with
\be
\hat G_B=\hat p_B\, t - m_B\,\hat x_B \,\,.
\ee

It is then straightforward to check that the following set of seven operators 
\be
\begin{array}{ll} \label{eq:dyngenerators}
 \hat P_{AB}= \hat x_A\otimes \hat p_B,  &\hat K_{AB}= \frac{\hat p_A}{m_A}\,\otimes \hat G_B, \\[3pt]
 \hat D_A= \frac{1}{2}\left(\hat x_A\,\hat p_A + \hat p_A\,\hat x_A \right)\,\otimes \mathds{1}_B, & \hat D_B= \mathds{1}_A \, \otimes \frac{1}{2} \left(\hat x_B \, \hat p_B + \hat p_B \, \hat x_B \right), \\[3pt]
 \hat Q_A=\frac{\hat p_A^2}{2\,m_A} \otimes \mathds{1}_B, & \hat Q_B = \mathds{1}_A \otimes \frac{\hat p_B^2}{2\,m_B}, \qquad\qquad \hat T=\hat p_A\otimes \hat p_B,
\end{array}
\ee
closes again a Lie algebra, whose commutation rules are given by
\begin{equation} 
	\begin{aligned}
		&\multispan3{$\big[\hat P_{AB}, \hat K_{AB}\big]=  i \hbar\, \dfrac{m_B}{m_A}\,\hat D_A-i \kappa\, \dfrac{m_B}{m_A}\,\hat D_B +  2i\kappa \dfrac{m_B}{m_A}\hat{Q}_B t$, \hfill} &&[\hat P_{AB}, \hat D_A]=\,i\,\kappa \,\hat P_{AB},\\
		 &[\hat P_{AB}, \hat D_B]=\,-i\,\hbar \,\hat P_{AB},
		&&[\hat P_{AB}, \hat Q_A]=  i \frac{\kappa}{m_A} \, \hat T, &&[\hat P_{AB}, \hat Q_B]=  0,\\
		  &[\hat P_{AB}, \hat T]= 2\,i\,\kappa\,m_B\,\hat Q_B, &&[\hat K_{AB}, \hat D_A]= \,-i\,\kappa \,\hat K_{AB}, &&[\hat K_{AB}, \hat D_B]= \,i\,\hbar \,\hat K_{AB} \,-2\,i \frac{\hbar}{m_A} \, \hat T\, t,\\
		 &[\hat K_{AB}, \hat Q_A]= 0,  &&[\hat K_{AB}, \hat Q_B]= \,  -i \frac{\hbar}{m_A} \, \hat T, &&[\hat K_{AB}, \hat T]=  -2\,i\,\hbar\,m_B\,\hat Q_A,\\
	   &[\hat D_A, \hat Q_A]=  2\,i\,\kappa\, \hat Q_A, &&[\hat D_A,\hat D_B]= 0, &&[\hat D_A, \hat Q_B]=  0,\\
	   &[\hat D_A, \hat T]=  \,i\,\kappa\, \hat T, &&[\hat D_B, \hat Q_B]=  2\,i\,\hbar\, \hat Q_B, &&[\hat D_B, \hat Q_A]=  0,\\ 
	 &[\hat D_B, \hat T]=  \,i\,\hbar\, \hat T.
	\end{aligned}
	 \label{eq:dynamicalalgebra}
\end{equation}
We call this algebra the {\it dynamical Lie algebra} $\mathcal{D}(7)$ of QRF transformations, which is actually a one-parametric family of Lie algebras with parameter $t$. Note that the fact that all the generators of the dynamical Lie algebra given by Eq.~\eqref{eq:dyngenerators} are quadratic functions in terms of the phase space operators is essential in order to guarantee that they close a Lie algebra.

We recall that in the previous section we were forced to introduce the generators $\hat D_{A}$ and $\hat D_{B}$ in order to close the algebra containing $\hat P_{AB}$ and $\hat K_{AB}$. Then it turned out that these generators were exactly the ones we needed in order to build the extended symmetry transformations of Sec.~\ref{sec:IntroQRF} in the $t=0$ case. Analogously, now we see that when $t\neq0$ the set $\lbrace \hat P_{AB}, \hat K_{AB}, \hat D_A, \hat D_B \rbrace$ is no longer a subalgebra. The requirement that these generators close some larger algebra  when $t\neq 0$ leads us to introduce further operators $\lbrace \hat Q_{A}, \hat Q_{B}, \hat T\rbrace$, which are indeed needed to build the extended symmetry  transformations $\hat S_{T}\,, \hat S_{b}$ of the previous section in the $t\neq 0$ case, Eqs.~\eqref{eq:extT} and \eqref{eq:extb}. 

So, as already stated for the non-dynamical algebra of Sec.~\ref{sec:FrozenAlg}, the generators of the algebra of Eq.~\eqref{eq:dynamicalalgebra} constitute the building blocks of the extended symmetry transformations $\hat S_{T}$ and $\hat S_{b}$. The only element that is missing to write the transformations $\hat S_{T}$ and $\hat S_{b}$ is the parity-swap operator $\mathcal{\hat{P}}_{AC}$. However, the algebraic structure can be determined independently of the parity-swap operator. The role of $\mathcal{\hat{P}}_{AC}$ is to ensure that the full set of relational variables is mapped consistently, but it does not influence the group structure of the QRF transformations. This fact allows us to identify the physical meaning of the additional generators  $\lbrace \hat Q_{A}, \hat Q_{B}, \hat T\rbrace$ appearing in Eq.~\eqref{eq:dynamicalalgebra}. Specifically, the operators $\hat Q_{A}, \hat Q_{B}$ generate the free motion of the particles $A$ and $B$, respectively. In the extended symmetry transformation framework this kind of operators is introduced in order to enforce the invariance of the  Hamiltonian of the quantum system. The action of the operator $\hat T$ on the position and momentum operators is
\begin{equation}
	\begin{split}
		&e^{\frac{i}{\hbar}\alpha \hat{T}} \hat{x}_A e^{-\frac{i}{\hbar}\alpha \hat{T}} = \hat{x}_A +\frac{\kappa}{\hbar} \alpha \hat{p}_B, \qquad e^{\frac{i}{\hbar}\alpha \hat{T}} \hat{p}_A e^{-\frac{i}{\hbar}\alpha \hat{T}} = \hat{p}_A,\\
		&e^{\frac{i}{\hbar}\beta \hat{T}} \hat{x}_B e^{-\frac{i}{\hbar}\beta \hat{T}} = \hat{x}_B + \beta \hat{p}_A, \qquad e^{\frac{i}{\hbar}\beta \hat{T}} \hat{p}_B e^{-\frac{i}{\hbar}\beta \hat{T}} = \hat{p}_B,
	\end{split}
\end{equation}
where $\alpha$ and $\beta$ have the physical dimensions of a time divided by a mass. Hence, the physical meaning of $\hat{T}$ is to generalise to QRFs the Galilean transformation $x' = x - vt$ without necessarily transforming the velocity through the momentum $\hat{p}_B$. Notice that the operator $\hat{T}$ appears in the $\hat{S}_b$ transformation, when the time-dependent Galilean boost is explicitly written as $\hat{U}_{\hat{K}}= e^{\frac{i}{\hbar}\frac{\hat{p}_A}{m_A}\hat{G}_B}$ and $\hat{G}_B = \hat{p}_B t - m_B \hat{x}_B$.

From the point of view of the four-dimensional quantum phase space $(\hat x_A,\hat p_A,\hat x_B,\hat p_B)$,  each of the generators $\hat X$ of the dynamical Lie algebra provides an action on such phase space through the corresponding group element
\begin{equation}
\hat U_{\hat X}={\rm e}^{i \lambda \hat X},
\end{equation}
with some appropriate (commutative) transformation parameter $\lambda$.
By explicitly computing the action of the group element $\hat U_{\hat X}$ for any $\hat X$ of the algebra of Eq.~\eqref{eq:dynamicalalgebra} on the phase space coordinates it is easy to see that $\hat U_{\hat X}$ generates a linear {\it canonical transformation}, see Table \ref{table1}, as it was expected since $\hat U_{\hat X}$ is always a unitary operator. Note that the group of linear canonical transformations in a $2N$-dimensional phase space is the real symplectic group $Sp(2N,\mathbb R)$~\cite{MQjmp71}. For $N=2$ we have that $Sp(4,\mathbb R)$ is 10-dimensional,  and the dynamical Lie group generated by the Lie algebra~\eqref{eq:dynamicalalgebra} is a 7-dimensional subalgebra of $Sp(4,\mathbb R)$.

\begin{table}[h]
 \begin{center}
\noindent
\begin{tabular}{l l l l l }
\hline
\hline
\\[-0.2cm] 
  & $\hat x_A$ &  $\hat p_A$ &  $\hat x_B$ &  $\hat p_B$   \\[0.2cm]
\hline
\hline
\\[-0.2cm]
$U_P={\rm e}^{\frac{i}{\hbar} \hat P_{AB}}$  & $\hat x_A$ &  $\hat p_A - \frac{\kappa}{\hbar}\, \hat{p}_B$ &  $\hat x_B + \hat{x}_A$ &  $\hat p_B$   \\[0.2cm]
\hline
\\[-0.2cm]
$U_G={\rm e}^{\frac{i}{\hbar}  \hat K_{AB}}$  & $\hat x_A + \frac{\kappa}{\hbar}\, \frac{1}{m_A}\, (\hat{p}_B t -m_B \hat{x}_B)$ &  $\hat p_A$ &  $\hat x_B+ t\, \frac{\hat{p}_A}{m_A}$ &  $\hat p_B+ \frac{m_B}{m_A}\,\hat{p}_A$   \\[0.2cm]
\hline
\\[-0.2cm]
${\rm e}^{\frac{i}{\kappa}\alpha \hat D_A}$  & ${\rm e}^{ \alpha} \hat x_A$ &  ${\rm e}^{- \alpha} \hat p_A$ &  $\hat x_B$ &  $\hat p_B$   \\[0.2cm]
\hline
\\[-0.2cm]
${\rm e}^{\frac{i}{\hbar}\alpha \hat D_B}$  & $\hat x_A$ &  $\hat p_A$ &  ${\rm e}^{\beta} \hat x_B$ &  $ {\rm e}^{-\beta} \hat p_B$   \\[0.2cm]
\hline
\\[-0.2cm]
${\rm e}^{\frac{i}{\kappa}\alpha \hat Q_A}$  & $\hat x_A + \frac{\alpha}{m_A}\,\hat p_A$ &  $\hat p_A$ &  $\hat x_B$ &  $\hat p_B$   \\[0.2cm]
\hline
\\[-0.2cm]
${\rm e}^{\frac{i}{\hbar}\alpha \hat Q_B}$  & $\hat x_A$ &  $\hat p_A$ &  $\hat x_B+ \frac{\alpha}{m_B}\,\hat p_B$ &  $\hat p_B$   \\[0.2cm]
\hline
\\[-0.2cm]
${\rm e}^{\frac{i}{\hbar}\,\alpha \hat T}$ & $\hat x_A + \alpha\,\frac{\kappa}{\hbar}\,\hat p_B$ &  $\hat p_A$ &  $\hat x_B+ \alpha\,\hat p_A$ &  $\hat p_B$   \\[0.2cm]
\hline
\\[-0.8cm]
\end{tabular}
\end{center}
\caption{\label{table1} \small  Action of the seven one-parametric subgroups of the dynamical algebra $\mathcal{D}(7)$ onto the quantum phase space variables for two generic particles, giving rise to the
corresponding canonical transformations. These uniparametric subgroups together with the parity-swap operator $\mathcal{\hat{P}}$ provide the building blocks from which QRF transformations~\eqref{eq:extT} and~\eqref{eq:extb} can be constructed. Notice that for $U_{P}$ and $U_{G}$ we have fixed the parameter $\lambda$ appearing in $U_{\hat X}=e^{i\lambda \hat X}$ to $\lambda=1/\hbar$, since this is the relevant case from the QRF transformations point of view. For the other generators we have assumed the parameter $\lambda$ to be proportional to the most appropriate between $1/\kappa$ and $1/\hbar$ in order to make contact with the exponential operators arising in the definition of QRF transformations, but we have left the proportionality constant  $\alpha$ free in order to consider more general phase space transformations than the ones related to the extended symmetry transformations of Sec.~\ref{sec:IntroQRF}.}
\end{table} 


Finally, we stress that if we specialize the full dynamical algebra~\eqref{eq:dynamicalalgebra} when $t=0$, and we define the new generator
\begin{equation}
	\hat D= \kappa\, \frac{\mathds{1}_A}{m_A}\otimes m_B \hat D_B - 
\hbar \, \frac{\hat D_A}{ m_A}\,\otimes m_B \mathds{1}_B \, ,
\end{equation}
then the operators $\lbrace \hat P_{AB}, \hat K_{AB}, \hat D, \hat Q_A, \hat Q_B, \hat T \rbrace$ generate a 6D subalgebra where $\lbrace \hat P_{AB}, \hat K_{AB}, \hat D \rbrace$ define a $su(1,1)\simeq sl(2,\mathbb R)$ algebra and $\lbrace \hat Q_A, \hat Q_B, \hat T \rbrace$ is a 3D Abelian sector. As it is explicitly discussed in  Appendix \ref{App:Poincare}, this 6D algebra turns out to be isomorphic to the Poincar\'e algebra of special relativistic transformations in (2+1) dimensions, and in this way an ``accidental'' dynamical Poincar\'e symmetry arises in the context of QRF transformations at $t=0$. Once again, this fact shows that the introduction of noncommutative Galilean translation and boost parameters changes completely the algebraic framework for the theory in which inertial transformations for QRFs have to be defined.


\subsect{Composition of two QRF transformations}
\label{sec:GroupLaw}

We are now in the position of answering the question we asked at the end of Sec.~\ref{sec:IntroQRF}, and write the transformation obtained by composing two QRF transformations as a function of the QRF group elements. In particular, we  look for an explicit expression of the transformation $\hat{S}_{D}^{(C \rightarrow A)} \equiv \hat{S}_b^{(B \rightarrow A)} \hat{S}_T^{(C \rightarrow B)} $, \emph{i.e.},
\begin{equation}
\hat{S}_{D}^{(C \rightarrow A)} = e^{-\frac{i}{\hbar}\hat{Q}_B^{(A)} t} \mathcal{\hat{P}}_{AB} e^{\frac{i}{\hbar} \ln \left(\frac{m_B}{m_A}\right) \hat{D}_A^{(B)}} e^{\frac{i}{\hbar} \hat{K}_{AC}^{(B)}}  e^{\frac{i}{\hbar}\hat{Q}_A^{(B)} t}        e^{-\frac{i}{\hbar}\hat{Q}_C^{(B)} t} \mathcal{\hat{P}}_{BC} e^{\frac{i}{\hbar} \hat{P}_{BA}^{(C)}}  e^{\frac{i}{\hbar}\hat{Q}_B^{(C)} t},
\end{equation}
 where as discussed in Sec.~\ref{sec:IntroQRF} we have set $\kappa = \hbar$. The action on the phase space coordinates is the following
\begin{equation}
	\begin{split}
		&\hat{S}_D \hat{x}_A^{(C)} \hat{S}_D^\dagger = -\frac{m_B}{m_A}\hat{x}_B^{(A)} - \frac{m_A + m_C}{m_A}\hat{x}_C^{(A)} +\hat{p}_B^{(A)} t \left[ \frac{1}{m_A}-\frac{1}{m_{B}}\right] + \hat{p}_C^{(A)}t\left[\frac{1}{m_A}+ \frac{1}{m_C}\right],\\
		&\hat{S}_D \hat{p}_A^{(C)} \hat{S}_D^\dagger =  - \frac{m_A}{m_B}\hat{p}_B^{(A)},\\
		&\hat{S}_D \hat{x}_B^{(C)} \hat{S}_D^\dagger = - \hat{x}_C^{(A)} + \hat{p}_C^{(A)} t\left[\frac{1}{m_C} - \frac{1}{m_B}\right] + \frac{m_C + m_A}{m_B}\frac{\hat{p}_B^{(A)}}{m_B}t,\\
		&\hat{S}_D \hat{p}_B^{(C)} \hat{S}_D^\dagger = -\hat{p}_C^{(A)} + \frac{m_A + m_C}{m_B}\hat{p}_B^{(A)}.
	\end{split}
\end{equation} 
A lengthy computation shows that this new transformation can be written as a QRF transformation from $C$ to $A$ in terms of the generators of Eq.~\eqref{eq:dyngenerators} as follows:
\begin{equation}
	\begin{split}
		& \hat{S}_D =  \mathcal{\hat{P}}_{AC} \, e^{-\frac{i}{\hbar}\frac{m_A}{m_C}\hat{Q}_A^{(C)} t} \, e^{-\frac{i}{\hbar}\left(1 - \frac{m_C^2}{m_B^2}\right)\hat{Q}_B^{(C)} t} \, e^{\frac{i}{\hbar}\ln\left(\frac{m_C}{m_A}\right) \hat{D}_A^{(C)} } \, e^{\frac{i}{\hbar}\ln\left(\frac{m_B}{m_C}\right) \hat{D}_B^{(C)} }  \\
		& \qquad\qquad\qquad \times e^{-\frac{i}{\hbar}\left(\frac{m_A}{m_C}\right)^2\hat{P}_{AB}^{(C)}} \, e^{\frac{i}{\hbar}\frac{m_C}{m_B}\hat{K}_{AB}^{(C)}} \, e^{\frac{i}{\hbar}\frac{m_A}{m_C}\hat{P}_{AB}^{(C)}}\, e^{\frac{i}{\hbar}\hat{Q}_A^{(C)} t}.
			\end{split}
\end{equation}

The transformation $\hat{S}_D$ can also be shown to be an extended symmetry of the free Hamiltonian $\hat{H}_{AB}^{(C)} = \frac{(\hat{p}_A^{(C)})^2 }{2m_A}+ \frac{(\hat{p}_B^{(C)})^2}{2m_B}$, because
\begin{equation}
	\hat{H}_{BC}^{(A)} = \hat{S}_D \hat{H}_{AB}^{(C)} \hat{S}_D^\dagger + i\hbar \frac{d \hat{S}_D}{dt} \hat{S}_D^\dagger =  \frac{\left(\hat{p}^{(A)}_B\right)^2}{2m_B} + \frac{\left(\hat{p}^{(A)}_C\right)^2}{2m_C}.
\end{equation}

Some intuition concerning the roots of the computational complexity of the previous results can be obtained by resorting to the well-known Baker-Campbell-Haussdorf (BCH) formula, which gives the algebraic background for Lie group multiplication formulae by providing the solution in $\hat Z$ for the equation
\be
e^{\hat X}\,e^{\hat Y} = e^{\hat Z} \, ,
\ee
where both $\hat X$ and $\hat Y$ belong to a Lie algebra $g$. Such solution can be explicitly given (see~\cite{Jacobson}) as a formal series in the Lie algebra elements whose first terms are given by
\be
\hat Z=\hat  X +\hat  Y + \frac12\,[\hat X,\hat Y] + \frac{1}{12}([\hat X,[\hat X,\hat Y]] - [\hat Y,[\hat X,\hat Y]]) \, ,
\ee
plus, in general, an infinite number of terms involving higher order commutators between $X$ and $Y$. In the particular case that $[\hat X,[\hat X,\hat Y]] = [\hat Y,[\hat X,\hat Y]]=0$, all higher order terms vanish and we have that
\be
e^{\hat X}\,e^{\hat Y} = e^{\hat X +\hat  Y + \frac12\,[\hat X,\hat Y]} \, .
\ee
This is exactly what happens when we consider the multiplication of a standard Galilean boost generated by $\hat G\equiv \hat X$ and a Galilean translation generated by $\hat P\equiv \hat Y$, since $[\hat G, \hat P] =i\hbar\, \hat M$ and the mass generator commutes with both $\hat G$ and $\hat P$, see Eq.~\eqref{gal11}. Therefore, if we consecutively apply a Galilean boost and a translation to a one-particle quantum state, the only difference that we obtain with respect to applying the same transformations in reverse ordering is a phase factor which depends on the mass of the particle.
However, if we are dealing with the generalized translation and boost transformations~\eqref{tyb} for QRFs, the situation is much more involved. Now, the BCH series no longer finishes with the first commutator $[\hat K_{AB},\hat P_{AB}]$, and the full series has to be computed by including  the new Lie algebra generators that arise as higher order commutators. In fact, the essential result that we have shown is that this BCH multiplication can be self-consistently closed for all the basic QRF transformations if we consider the 7-dimensional dynamical Lie group defined by Eq.~\eqref{eq:dynamicalalgebra}.

\sect{The classical reference frame limit}
\label{sec:Classical}

It is well-known that the centrally extended Galilei group is the symmetry group of the rays of Hilbert space solutions of the Schr\"odinger equation for a massive particle~\cite{levy1971galilei,LevyLeblond:1974zj}. We recall that the (1+1) centrally extended Galilei algebra generating such group of inertial transformations for a quantum particle is given by
\be
[{\hat G},\hat P_0]=i\,\hbar\,\hat P, \qquad
[{\hat G},\hat P]= i\,\hbar\,\hat M, \qquad
[\hat P_0,\hat P]=0,\qquad
[\hat M,\cdot]=0,
\label{gal11}
\ee
where ${\hat G},\hat P,\hat P_0$ are, respectively, the generators of boost transformations, space translations and time translations, while  $\hat M$ is the central extension (mass operator) which has to be added to the usual Galilei transformations in order to recover the invariance (up to a phase) of the Schrödinger wave functions.
The Casimir operator for this algebra is
\be
\hat C=2\,\hat M\,\hat P_0-\hat P^2 \,,
\ee
from which the one-particle Hamiltonian of a free Galilean particle in (1+1) dimensions can be identified with the time-translation generator
\be
\hat P_0=\frac{\hat P^2}{2\,\hat M} + \hat C \, ,
\ee
as expected. The one-particle ($B$) representation of the Galilei algebra with $\hat C=0$ is given by
\be
\hat P_0=\frac{\hat p_B^2}{2\,m_B}\, ,
\qquad
{\hat G= \hat p_B\, t - m_B\,\hat x_B} \, ,
\qquad
\hat P=- \hat p_B\, ,
\qquad
\hat M= m_B \, .
\label{repG}
\ee
Therefore, a (1+1) centrally extended Galilei inertial transformation is given by the group element
\be
g={\rm e}^{\frac{i}{\hbar}\,\theta\,\hat M}\,{\rm e}^{\frac{i}{\hbar}\,b\,\hat P_0}\,{\rm e}^{\frac{i}{\hbar}\,a\,\hat P}\,{\rm e}^{\frac{i}{\hbar}\,v\,{\hat G}}\,,
\ee
where a Galilei group element is parametrized as $(\theta,b,a,v)$.

In our framework, the classical reference frame limit of the generators of the dynamical Lie algebra defined in Eq.~\eqref{eq:dynamicalalgebra} is algebraically given by taking the $\kappa\to 0$ in the algebra, which at the level of the representation in Eq.~\eqref{eq:dyngenerators} is tantamount to saying that the positions and momenta for particle $A$ become commutative functions, that we will denote as objects without hats $(\hat x_A\rightarrow x_A,\hat p_A\rightarrow p_A)$. Note that the separate existence of the parameter $\kappa$  allows us to consider formally this limit as different from the usual classical limit $\hbar\to 0$, since we do not want to suppress the quantum nature of particle $B$. This is why we call this procedure a ``classical reference frame limit'', {under which the generators $\{\hat P_{AB},\hat K_{AB}, \hat Q_B \}$   of the dynamical algebra~\eqref{eq:dyngenerators} have the following classical counterparts
\begin{equation} \label{crfalgebra} 
	\begin{split}
		& \hat P_{AB}^c=  x_A\otimes \hat p_B, \\
		&  \hat K_{AB}^c= \frac{p_A}{m_A}\,\otimes (\hat p_B\, t - m_B\,\hat x_B), \\
		&   \hat Q_B^c= \mathds{1}_A \otimes \frac{\hat p_B^2}{2\,m_B}\, ,
	\end{split}
\end{equation}
where $x_A$ and $p_A$ are now commuting functions, which act as constant operators on the left hand side of the tensor product. The commutation rules among the operators of Eq.~\eqref{crfalgebra} are just the $\kappa\to 0$ limit of~\eqref{eq:dyngenerators}, and it is straightforward to check that if we define
\be
\hat P_0=\hat Q^c_B
\qquad
{\hat G}= \frac{m_A}{p_A}\, \hat K_{AB}^c
\qquad
\hat P=-\frac{1}{x_A}\,\hat P_{AB}^c
\qquad
\hat M=m_B \, {\mathds{1}_B},
\ee
we obtain the representation~\eqref{repG} for the (1+1) centrally extended Galilei Lie algebra~\eqref{gal11}. Therefore, when $\kappa\neq 0$ the extended Galilei algebra is by no means a subalgebra of the dynamical Lie algebra $\mathcal{D}(7)$, but it can be obtained as a subalgebra of its $\kappa\rightarrow 0$ limit under a suitable redefinition of the generators that takes into account the values of the phase space variables for particle $A$.

Moreover, this classical reference frame limit gives rise to the classical symplectic structure on the (now commutative) algebra of functions on the phase space for particle $A$ through the standard definition of the Poisson bracket as a limit of the commutator, namely
\be
\{ x_A, p_A\}=\lim_{\kappa\to 0}\frac{[\hat x_A,\hat p_A]}{i\,\kappa} =1\, .
\ee
In this sense, the $\kappa\to 0$ limit gives rise to the classical Hamiltonian structure for the $A$ variables, which can still be considered as dynamical ones from a classical mechanics perspective. Therefore, the $\kappa\to 0$ limit provides the appropriate tool in order to obtain classical dynamical reference frames, where the non-vanishing commutation rule between the quantum boost and translation parameters becomes a non-vanishing canonical Poisson bracket between their classical counterparts.

We stress that this classical reference frame limiting procedure can be considered algebraically analogous to the one providing the classical electromagnetic field description of a quantum matter system {$B$} interacting with a {single-mode} quantum electromagnetic field {$A$, which we describe as a quantum harmonic oscillator}. In this case, the full interacting system {is} described {as a state living on} the tensor product Hilbert space {$\mathcal{H}= \mathcal{H}_A \otimes \mathcal{H}_B$. Here, $\mathcal{H}_A$ is the Hilbert space} of  the quantum harmonic oscillator, where the quadrature field operators satisfy the canonical commutation rule $[\hat x_A,\hat p_A]=i\,\hbar$, {and $\mathcal{H}_B$ is} the Hilbert space encoding the quantum matter. In the case that the quantum system $B$ can be effectively described as a two-level system, the algebra of observables for $B$ is the $su(2)$ Lie algebra, while the {electromagnetic} field algebra is the Heisenberg-Weyl algebra (thus giving rise to the so-called Dicke model~\cite{Dicke}). Then the classical electromagnetic field description of the system can be obtained by performing the ``strong field'' limit, which arises when the harmonic oscillator states $A$ of the quantum field have very high quantum numbers {or, equivalently, that the number of photons is much larger than the number of atoms in the sample~\cite{Chumakov}}. This limit would be algebraically equivalent to taking the $\hbar\to 0$ limit {\em  for the A Hilbert space only}, while the quantum system $B$ remains untouched. Indeed, the algebraic structure and the representation theory underlying the quantum field -- quantum matter model are much more involved than the corresponding ones for the classical field - quantum matter one {(see~\cite{Ballesteros_1999})}.



\section{Comments and conclusions}

In this work we have contributed to the understanding of the role of quantum reference frame (QRF) transformations  as symmetries of quantum mechanical systems. It had been previously demonstrated that QRF transformations are extended symmetries of the free particle Hamiltonian, in the sense explained around Eq.~\eqref{eq:ExtSymm}. However, an important property of symmetry transformations is that they should close a group structure, but this was still unproven. We found that the difficulties encountered so far could be traced back to the fact that the group structure is not apparent at the level of the full QRF transformations, but it is realised at the level of their individual building blocks defining the canonical transformations on the phase space of the quantum system which includes the QRFs. They consist of the noncommutative generalisation of Galilean translation and boost generators, together with other operators that were  included in the QRF transformations in order to ensure they preserved the relational nature of the phase space coordinates and that they were extended symmetries of the Hamiltonian, namely velocity-rescaling operators and time-evolution operators. What is interesting is that, even though we started from just the generalised translation and boost generators, we were led to add the other transformations when searching for the Lie algebra they could close.  The Lie algebra we obtained is a 7-dimensional Lie algebra, which recovers  the usual centrally extended Galilean algebra as a subalgebra when the classical reference frame limit is taken. Thanks to the identification of the elements which close the group structure, we were able to show that indeed the composition of two QRF transformations can be written in terms of these elements, so that their role as symmetries of the quantum system is reinforced. Also, the introduction of a second parameter $\kappa$ governing the noncommutativity of the operators appearing in the  QRF transformations makes it possible to consider a classical reference frame limit of the algebra of symmetries, under which we recover the usual Galilean symmetries from a dynamical framework.

From a more technical viewpoint, in Sec.~\ref{sec:TimeDepAlg} we stressed the relevance of linear canonical transformations in the definition of the symmetry algebra for quantum reference frames. This suggests that the symplectic group $Sp(2N,\mathbb R)$, as the group of linear canonical transformations on the $2N$-dimensional phase space, should play an outstanding role in the QRF description of  a system of $N$ quantum particles in (1+1) dimensions. In the particular setting for $N=2$  here presented, since the $sp(4,\mathbb R)$ Lie algebra is isomorphic to the Anti-de Sitter Lie algebra $so(3,2)$, the emergence of dynamical Lorentz symmetries in the context of Galilean QRFs, like the $so(2,1)$ one generated by the relational Lie algebra $\lbrace \hat P_{AB}, \hat K_{AB}, \hat D \rbrace$, becomes natural.} {In this sense, a deeper understanding of the representation theory of the full 7-dimensional dynamical Lie algebra is worth being explored, as well as the derivation of its complete group law for arbitrary group parameters. This could be helpful in order to understand in group-theoretical terms why in the QRF scenario we are constrained to choose a specific set of dynamical group parameters in order to define the specific symmetries of the system, while more generic group transformations would be possible.}

{Evidently, the generalization of the approach here presented to QRFs defined by Galilean particles living in (2+1) and (3+1) dimensions has to be faced in the near future. In this sense, preliminary studies show that the definition of noncommutative translations and boosts for additional dimensions can be done just by mimicking the approach here presented. Nevertheless, the introduction of rotation transformations for QRFs is by no means a trivial task, since the underlying issue of the definition of the noncommutative angle operator associated to a given rotation has to be fixed in a consistent way with respect to the rest of symmetries, a problem to which the literature on QRFs has already devoted some effort (see Ref.~\cite{perspective2}). In any case, we think that being able to set the symmetry problem for QRF transformations in a Lie group framework makes it possible to use all the well-known group-theoretical machinery in order to face this and other relevant open questions  from a novel perspective.}


\section*{Acknowledgements}

We would like to thank the anonymous Referees for their helpful suggestions and comments. A.B. acknowledges partial support by Ministerio de Ciencia e Innovaci\'on (Spain) under grants MTM2016-79639-P (AEI/FEDER, UE) and PID2019-106802GB-I00/AEI/10.13039/501100011033, and by Junta de Castilla y Le\'on (Spain) under grants BU229P18 and BU091G19. F.G. acknowledges support from Perimeter Institute for Theoretical Physics. Research at Perimeter Institute is supported in part by the Government of Canada through the Department of Innovation, Science and Economic Development and by the Province of Ontario through the Ministry of Colleges and Universities. The authors would like to acknowledge contribution from the COST Action CA18108.

\section*{Appendix}
\appendix

\section{Calculation of the action of the operator $\hat{D}$}
\label{App:CalcD}

\setcounter{equation}{0}
\renewcommand{\theequation}{A.\arabic{equation}}

Let us consider the operator $\hat{U}_D = e^{\frac{i}{\hbar}\alpha \hat{D}}$, where $\hat{D} = \frac{1}{2}(\hat{x}\hat{p} + \hat{p}\hat{x})$ (in the main text, the position and momentum operators can refer to both particles $A$ and $B$). Let us now consider the non symmetrised version of the operator $\hat{U}_D$, which we denote as $\hat{T}_D = e^{\frac{i}{\hbar} \alpha \hat{x} \hat{p}} $, which has the same action on the phase space operators as its symmetrised version, i.e., $\hat{U}_D \hat{x} \hat{U}_D^\dagger = \hat{T}_D \hat{x} \hat{T}_D^{-1}$, because $ \hat{U}_D = \hat{T}_D e^{\frac{1}{2}\alpha}$ and $ \hat{U}_D^\dagger = \hat{T}_D^{-1} e^{-\frac{1}{2}\alpha}$. Hence, we can write
\begin{equation}
	\hat{T}_D \hat{x} \hat{T}_D^{-1} = \sum_{n=0}^\infty \frac{1}{n!} \left( \frac{i}{\hbar} \alpha \hat{x} \hat{p} \right)^n \hat{x} \hat{T}_D^{-1}
\end{equation}
If we define $\hat{A}= \frac{i}{\hbar} \alpha \hat{x} \hat{p}$  we find 
\begin{align}
	& \hat{A} \hat{x} = \hat{x} \hat{A} + \alpha \hat{x};\\
	& \hat{A}^2 \hat{x} = \hat{x} \hat{A}^2 + 2 \alpha \hat{x} \hat{A} + \alpha^2 \hat{x};\\
	& \hat{A}^3 \hat{x} = \hat{x} \hat{A}^3 + 3 \alpha \hat{x} \hat{A}^2 + 3 \alpha^2 \hat{x} \hat{A} + \alpha^3 \hat{x};\\
	& \qquad\vdots\\
	& \hat{A}^n \hat{x} = \sum_{m=0}^n \binom{n}{m}\left(\alpha^m \hat{x}\right) \hat{A}^{n-m} .
\end{align}
We can thus rewrite
\begin{equation}
	\begin{split}
		\hat{T}_D \hat{x} &= \sum_{n=0}^\infty \sum_{m=0}^n \frac{1}{m!(n-m)!}\left(\alpha^m \hat{x}\right) \hat{A}^{n-m}=\\
		&= \sum_{m=0}^\infty \sum_{\ell=0}^\infty \frac{1}{m!\ell!}\left(\alpha^m \hat{x}\right) \hat{A}^{\ell}=\\
	&= e^{\alpha} \hat{x} \hat{T}_D.
	\end{split}
\end{equation}
We then find
\begin{equation}
	\hat{U}_D \hat{x} \hat{U}_D^\dagger = e^{\alpha} \hat{x}, \qquad \hat{U}_D \hat{p} \hat{U}_D^\dagger = e^{-\alpha} \hat{p}.
\end{equation}
If we now define, e.g., $\alpha = \ln \frac{m_C}{m_A}$ and we look for the action of the operator $e^{\frac{i}{\hbar}\ln \frac{m_c}{m_A} \hat{D}_A}$ as defined in the main text, we find
\begin{equation}
	e^{\frac{i}{\hbar}\ln \frac{m_c}{m_A} \hat{D}_A} \hat{x}_A^{(C)} e^{-\frac{i}{\hbar}\ln \frac{m_C}{m_A} \hat{D}_A} = \frac{m_C}{m_A} \hat{x}_A^{(C)}, \qquad e^{\frac{i}{\hbar}\ln \frac{m_c}{m_A} \hat{D}_A} \hat{p}_A^{(C)} e^{-\frac{i}{\hbar}\ln \frac{m_C}{m_A} \hat{D}_A} = \frac{m_A}{m_C} \hat{p}_A^{(C)}.
\end{equation}
By also adding the parity-swap operator, we find the desired result
\begin{equation}
	\begin{split}
		& \hat{\mathcal{P}}_{AC} e^{\frac{i}{\hbar}\ln \frac{m_c}{m_A} \hat{D}_A} \hat{x}_A^{(C)} e^{-\frac{i}{\hbar}\ln \frac{m_C}{m_A} \hat{D}_A} \hat{\mathcal{P}}_{AC}^\dagger = -\frac{m_C}{m_A} \hat{x}_C^{(A)}, \\
		& \hat{\mathcal{P}}_{AC} e^{\frac{i}{\hbar}\ln \frac{m_c}{m_A} \hat{D}_A} \hat{p}_A^{(C)} e^{-\frac{i}{\hbar}\ln \frac{m_C}{m_A} \hat{D}_A} \hat{\mathcal{P}}_{AC}^\dagger = -\frac{m_A}{m_C} \hat{p}_C^{(A)}.
	\end{split}
\end{equation}

\section{(2+1) Poincar\'e symmetry arising at $t=0$}
\label{App:Poincare}
\setcounter{equation}{0}
\renewcommand{\theequation}{B.\arabic{equation}}

{As we mentioned in Section~\ref{sec:TimeDepAlg}, if we consider the commutation rules~\eqref{eq:dynamicalalgebra} when $t=0$, and we define the new generator}
\begin{equation}
	\hat D= \kappa\, \frac{\mathds{1}_A}{m_A}\otimes m_B \hat D_B - 
\hbar \, \frac{\hat D_A}{ m_A}\,\otimes m_B \mathds{1}_B \, ,
\end{equation}
then {a 6D Lie subalgebra is generated by} $\lbrace \hat P_{AB}, \hat K_{AB}, \hat D, \hat Q_A, \hat Q_B, \hat T \rbrace$, {with commutation rules given by}
\be
\begin{array}{lll} 
[\hat P_{AB},\hat K_{AB}]=-i\,\hat D , &\qquad 
[\hat P_{AB},\hat D]=-2\,i\,\kappa\,\hbar\,\frac{m_B}{m_A} \,\hat P_{AB}, &\qquad  
[\hat K_{AB},\hat D]=  2\,i\,\kappa\,\hbar\,\frac{m_B}{m_A} \,\hat K_{AB} , \\[2pt]
[\hat P_{AB}, \hat Q_A]=  i \frac{\kappa}{m_A} \, \hat T ,&\qquad
[\hat P_{AB}, \hat T]= 2\,i\,\kappa\,m_B\,\hat Q_B ,&\qquad
[\hat P_{AB}, \hat Q_B]=  0  , \\[2pt]
[\hat K_{AB}, \hat Q_A]= 0 ,&\qquad
[\hat K_{AB}, \hat Q_B]=   -i \frac{\hbar}{m_A} \, \hat T  ,&\qquad
[\hat K_{AB}, \hat T]=  - 2\,i\,\hbar\,m_B\,\hat Q_A  ,\\[2pt]
[\hat D, \hat Q_A]=  -2\,i\,\kappa\,\hbar\,\frac{m_B}{m_A}\,\hat Q_A ,&\qquad
[\hat D, \hat Q_B]=  2\,i\,\kappa\,\hbar\,\frac{m_B}{m_A}\,\hat Q_B ,&\qquad
[\hat D, \hat T]= 0 ,\\[2pt]
[\hat Q_A, \hat Q_B]=  0 ,&\qquad
[\hat Q_A, \hat T]=  0 ,&\qquad
[\hat Q_B, \hat T]= 0 \, .
\end{array}
\label{relalg} 
\ee
{Note that this algebra has a semidirect product structure where}  $\lbrace \hat P_{AB}, \hat K_{AB}, \hat D \rbrace$ define a $su(1,1)\simeq sl(2,\mathbb R)\simeq so(2,1)$ Lie subalgebra {(see also~\eqref{eq:RelationalAlgebra})} and $\lbrace \hat Q_A, \hat Q_B, \hat T \rbrace$ provide a 3D Abelian {sector}. Surprisingly enough,  it can be shown through a straightforward computation that this 6D Lie subalgebra
is isomorphic to the (2+1) Poincar\'e Lie algebra of relativistic inertial transformations 
\be
\begin{array}{lll} 
[\hat J,\hat P_i]= i\,  \epsilon_{ij}\hat P_j , &\qquad
[\hat J,\hat K_i]=  i\, \epsilon_{ij}\hat K_j , &\qquad  [\hat J,\hat P_0]= 0  , \\[2pt]
[\hat P_i,\hat K_j]=-i\,\delta_{ij}\hat P_0 ,&\qquad [\hat P_0,\hat K_i]=-i\,\hat P_i ,&\qquad
[\hat K_1,\hat K_2]= -i\,\hat J   , \\[2pt]
[\hat P_0,\hat P_i]=0 ,&\qquad [\hat P_1,\hat P_2]= 0  ,
\end{array}
\label{ba} 
\ee
where $i,j=1,2$, and  $\epsilon_{ij}$ is a skew-symmetric tensor with $\epsilon_{12}=1$. {Here}   $\hat J$ is the rotation generator, $\{\hat K_1,\hat K_2\}$ are the generators of {special relativistic} boost transformations and $\{\hat P_0,\hat P_1,\hat P_2 \}$ are, {respectively, the time and space} translation generators. The change of basis relating the two algebras is given by
\bea
&&  \hat J= \frac{1}{2\sqrt{\kappa\,\hbar}}(\hat P_{AB} + \frac{m_A}{m_B} \hat K_{AB}) , \cr
&&  \hat K_1=  \frac{1}{2\sqrt{\kappa\,\hbar}}(\hat P_{AB} - \frac{m_A}{m_B} \hat K_{AB}),\cr
&&  \hat K_2=  \frac{m_A}{2\,{\kappa\,\hbar}\, m_B} \, \hat D, \\
&&  \hat P_0 =\hbar\, m_A \, \hat Q_A + \kappa\, m_B \, \hat Q_B, \cr
&&  \hat P_1=   \sqrt{\kappa\,\hbar}\,\hat T  , \cr
&&  \hat P_2= - \hbar\, m_A \, \hat Q_A + \kappa\, m_B \, \hat Q_B\, ,\nonumber
\label{poincdyn}
\eea
{while the inverse change of basis reads}
\bea
&& \hat D= 2\,\kappa\,\hbar\,\frac{m_B}{m_A}\, \hat K_2, \cr
&& \hat K_{AB}= \sqrt{\kappa\,\hbar}\,\frac{m_B}{m_A}\, (  \hat J - \hat K_1  ) ,\cr
&& \hat P_{AB}= \sqrt{\kappa\,\hbar}\, (  \hat J + \hat K_1  ) , \label{inverse}\\
&& \hat T= \frac{1}{\sqrt{\kappa\,\hbar}}\,\hat P_1 ,\cr
&& \hat Q_A=  \frac{1}{2\,\hbar\,m_A}\,(\hat P_0 - \hat P_2), \cr
&&\hat Q_B= \frac{1}{2\,\kappa\,m_B}\,(\hat P_0 + \hat P_2).\nonumber
\eea

As a consequence, the (2+1) Poincar\'e algebra arises as an {`accidental' } dynamical symmetry for QRFs {since it only holds for $t=0$.} In fact, in our approach we are considering a specific representation of the Poincar\'e algebra given by
\bea
&&  \hat J=  \frac{1}{2\,\sqrt{\kappa\,\hbar}}\, (\hat x_A \otimes \hat p_B - \hat p_A \otimes \hat x_B) \, , \cr
&&  \hat K_1= \frac{1}{2\,\sqrt{\kappa\,\hbar}}\, (\hat x_A \otimes \hat p_B + \hat p_A \otimes \hat x_B) \, ,\cr
&&  \hat K_2= -\frac{1}{4\,\kappa} (\hat x_A\, \hat p_A +  \hat p_A\,  \hat x_A )\otimes \mathds{1}_B + 
\frac{1}{4\,\hbar} \mathds{1}_A\otimes (\hat x_B\, \hat p_B +  \hat p_B\,  \hat x_B )\, ,\cr
&&  \hat P_0= \frac{\hbar}{2} \hat p_A^2 \otimes \mathds{1}_B + 
\frac{\kappa}{2} \mathds{1}_A\otimes  \hat p_B^2\,  ,\label{rep21}\\
&&  \hat P_1=  \sqrt{\kappa\,\hbar}\, \, p_A\otimes p_B \, ,\cr
&&  \hat P_2=  -\frac{\hbar}{2} \hat p_A^2 \otimes \mathds{1}_B +
\frac{\kappa}{2} \mathds{1}_A\otimes  \hat p_B^2\, . \nonumber
\eea
We recall that the two quadratic Casimir operators for the (2+1) Poincar\'e algebra are given by
\be
{\cal \hat C}=\hat P_0^2-{\hat P_{1}}^2-{\hat P_{2}}^2, \qquad
{\cal \hat W}=-\hat J \hat P_0+\hat K_1 \hat P_2-\hat K_2 \hat P_1 \, ,
\nonumber
\ee
and  in the representation of Eq.~\eqref{rep21} we have that ${\cal \hat C}= 0$ and ${\cal \hat W}= 0$.
Therefore we have a massless and spinless representation of the Poincar\'e algebra since both the mass and the Pauli-Lubanski Casimir have zero eigenvalues.

Some comments could be in order:
\begin{itemize}

\item The transformation~\eqref{inverse} provides a relativistic interpretation of the 6D dynamical algebra for QRF at $t=0$, in terms of the generators of (2+1) Poincar\'e transformations, which act on the 4D quantum phase space of the particles AB through the representation~\eqref{rep21}. In particular, the generators $\hat P_{AB}$ and $\hat K_{AB}$ that were introduced as the QRF version of the Galilean translation and boost operators can be now interpreted as the superposition of a Poincar\'e rotation $J$ and the boost $K_1$, while the $\hat D$ generator is the $K_2$ boost. Moreover, from~\eqref{poincdyn} we see that the relativistic energy $P_0$ turns out to be a superposition of the QRF Hamiltonians $\hat Q_A$ and $\hat Q_B$ weighted by the noncommutativity parameters $\hbar$ and $\kappa$. {However, note that that the change of basis~\eqref{inverse} is well-defined provided that both $\hbar$ and $\kappa$ do not vanish}.

\item The classical reference frame limit $\kappa\to 0$ has nothing to do with the nonrelativistic $c\to\infty$ limit of the Poincar\'e algebra. Such a nonrelativistic limit is obtained in the kinematical basis {by applying} the automorphism
\be
P_i\to \frac 1 c\, P_i,\qquad  K_i\to \frac 1 c\, K_i \, , \qquad i=1,2 \, ,
\label{add}
\ee
{where $c$ is the speed of light and, afterwards, by taking the $c\to \infty$ limit of the algebra (see, for instance,~\cite{Ballesteros:2019mxi}). The transformation~\eqref{add},} when applied onto~\eqref{inverse} shows that the generators $\lbrace \hat P_{AB}, \hat K_{AB},  \hat Q_A, \hat Q_B \rbrace$ of the dynamical algebra cannot be homogeneously transformed in the limit $c\to\infty$, and thus the nonrelativistic limit cannot be defined {for the generators of  QRF transformations}.


\end{itemize}

\end{document}